\documentclass[a4paper,12pt]{article}
\usepackage{amsmath}
\usepackage{amssymb}
\usepackage{cite}
\newcommand{\ul}{\underline}
\newcommand{\longpage}{\enlargethispage*{1cm}}

\begin{document}

\begin{titlepage}
\renewcommand{\thefootnote}{\fnsymbol{footnote}}

\rightline{\large TUW--01--10} 
\vspace{1.5cm}

\begin{center}

\textbf{\Large On the Gauge-Independence of the S-Matrix}\\
\vspace{7ex}
  W.\ Kummer\footnotemark[1]
  \vspace{7ex}

  {Institut f\"ur
    Theoretische Physik \\ Technische Universit\"at Wien \\ Wiedner
    Hauptstrasse  8--10, A-1040 Wien, Austria}
  \vspace{2ex}

\footnotetext[1]{E-mail: \texttt{wkummer@tph.tuwien.ac.at}}
\end{center}
\vspace{7ex}
\begin{abstract}
The S-matrix is invariant with respect to the variation of 
any (global) parameter involved in the gauge fixing 
conditions, if that variation is accompanied by a certain 
redefinition of  the basis of polarization vectors. 
Renormalizability of the underlying gauge 
theory is not required. The proof is nonperturbative and,  
using the ``extended'' BRS transformation,  quite simple.  
\end{abstract}
\end{titlepage}

\section{Introduction}

After half a century of quantum gauge theory still some 
confusion seems to prevail regarding the difference between 
gauge-invariance and gauge-in\-de\-pend\-ence. A closely related 
difficulty concerns the concept of a ``physical 
observable''. For example, to this day gauge-invariance of a 
certain operator to some implies that it is an ``observable'', 
despite the work of many authors many years ago \cite{klu75} --- 
mainly in the context of deep-inelastic scattering of 
leptons  --- that renormalization introduces an 
unavoidable mixture with other (ghost-dependent) 
operators\footnote{A similar mechanism involving the 
gauge parameter-dependent operators works in the (ghostfree) 
axial and temporal gauges \cite{kon77}.} 
in the gauge fixed theory. The 
physical observables in that case, in a quite roundabout 
manner, become only one aspect of the gauge-invariant 
operators: only the anomalous dimensions of those operators 
enter as ingredients in the S-matrix element. Thus the 
gauge-independence of the latter is the central issue, a topic 
usually just taken for granted in the literature. I am not 
aware of any experimental situation in which --- from the 
point of view of  quantum {\em field} theory, for which 
ordinary quantum mechanics, generalizing classical  point 
particles,  
represents only a subdomain ! --- anything else than an  
S-matrix element provides the link between theory and 
experiment.\footnote{In my view the definition of ``observables'' 
should also include a prescription how to measure those 
quantities, e.g.\ the eigenvalues of hermitian operators in 
ordinary quantum mechanics. Take the total angular momentum of 
such a system: In order to measure it, one must let the system 
interact with some ``external field'' (here magnetic field) which 
is (weakly) coupled to that quantity. But then in the larger 
system, including that interaction, one again effectively has to 
calculate some S-matrix element, involving asymptotic initial and 
final states. In this sense the experimental preparation of the 
incoming system and the interactions with the detecting device 
are ``asymptotic''.}
\longpage

The literature on the gauge-independence of that 
quantity is very scarce. Since the proof proposed by Costa 
and Tonin \cite{costa75}, apart from adapting their  
argument to a proof in the axial gauge\cite{K76} and to 
superfields\cite{KMS} I am only aware of\cite{GG00} where a proof 
for scattering of fermions in the Standard Model was given which 
uses the  Nielsen identities\cite{N75}.
However, at the time when e.g.\ first 
computations in quantum gravity, involving full back reaction 
of the background (at least in the spherically reduced case 
\cite{KLV97}) have become available, and when the issue of an S-matrix 
element in gravity\cite{Vass95} 
becomes important when scattering of particles 
through an intermediate ``virtual black hole'' can be 
considered in a specific gauge \cite{GKV00} the publication of 
a simple and nevertheless 
complete proof, which is not restricted 
to gauge theories based upon Lie groups, seems to be 
useful.\footnote{I have been 
reminded a few times and especially recently \cite{greenb} that such a 
simple proof, which has formed for many years  part of my 
lectures on quantum field theory at the Vienna University of Technology, 
should be made available.}  
It uses the concept of extended BRS 
symmetry and the related Slavnov-Taylor identity for the 
generating function of Green functions \cite{pig84} (Section 2), 
completely avoiding the detour\cite{N75} through Nielsen 
identities \cite{KLV97} which are identities determining the 
gauge-dependence of one-particle-irreducible (1pi) vertices. 
Generically such identities are more important for questions 
related to renormalization and to the gauge dependence of 
anomalies\cite{WKA}, although they have shown their usefulness 
for the definition of mass, width of unstable particles, 
effective actions etc.\ \cite{C-others}.

The ``physical'' poles of the propagator yield the 
definition of the polarizations to be used in the S-matrix 
(Section 3). We assume the existence of such poles for 
massive particles. Gauge fields can be included after 
regularization by spontaneous symmetry breaking, if that 
mechanism is not provided by Nature anyhow (as for the W- 
and Z-bosons). \\
Our proof is outlined in Section 4, whereas possible 
limitations and 
extensions are discussed in the Conclusions (Section 5).

\section{Extended Identity for Green Functions}

The argument will be based on the generating functional Z 
for Green functions which can be written as 

\begin{equation}
Z^{(0)}\, (j,k) = \int (d\, \phi_A) \, \exp i L^{(0)}
%eqn:1
\end{equation}

with the action (summation of indices includes integration)

\begin{equation}
L^{(0)} = L_{inv} + L^{(0)}_{gf} + j_A\, \phi_A + k_A 
(s\,\phi_A)\; .
%eqn:2
\end{equation}

The symbol $\phi_A = (\phi_i, A_\alpha)$  comprises gauge 
fields, matter fields $(\phi_i)$  and auxiliary fields $(A_\alpha)$, 
appearing as a 
consequence of the gauge fixing procedure. Because of the 
possible presence of fermions and of Faddeev-Popov (FP) ghosts  
grading  should be taken into account: 

\begin{equation}
\phi_A\, \phi_B = (-1)^{\rm AB} \, \phi_B\, \phi_A
%eqn:3
\end{equation}

In the exponent of $(-1)$ for a (anti-)commuting field 
$\phi_A$ we set $(A = 1)$ $A=0$. The bosonic nature of $L^{(0)}$ 
entails the same relation for the sources $j_A$. Gauge 
fixing leaves (2) invariant under the BRS-transformation 
\cite{becc74} 
\begin{equation}
\delta\, \phi_A = \delta\,\lambda\, s\, \phi_A\; ,
%eqn:4
\end{equation}
where $s^2 = 0$ on all $\phi_A$ and $\delta\lambda$ is a 
(global) anticommuting parameter. The explicit form of 
$s\phi_A$ need not be specified. According to Zinn-Justin 
\cite{zj75} sources $k_A$ for $s\phi_A$ have been introduced. Clearly 
$s\phi_A$ and thus $k_A$ have grading $A+1$. For the gauge 
fixing action $L^{(0)}_{gf}$ we only need that it is 
BRS-exact
\begin{equation}
L^{(0)}_{gf} = s\,\Psi\,(\phi,p)\, ,
%eqn:5
\end{equation}
where the (anticommuting) functional $\Psi$ depends by the 
gauge-fixing condition on gauge parameters which we take to 
be commuting (global) variables. It is sufficient to select 
one of them, called $p$, in what follows.\\
To give a concrete example, for the special case of linear 
inhomogeneous gauge fixing we may choose 
\begin{equation}
\Psi = \bar{b}_\alpha \, F_{\alpha i}\, \phi_i - 
\frac{\beta}{2}\, \bar{b}_\alpha \, B_\alpha \; ,
%eqn:6
\end{equation}
where $\bar{b}_\alpha$ is the antighost, related to the 
Nakanishi-Lautrup field $B_\alpha$ by $B_\alpha = s\, 
\bar{b}_\alpha, \; sB_\alpha = 0$. With $s \phi_i = 
D_{i\alpha} (\phi)\, c_\alpha$ containing the ghost-field 
$c_\alpha$ and $D_{i\alpha}$ from the gauge transformation 
$\delta \phi = D_{i\alpha}\, \delta \omega_\alpha$. Eqs.\ 
(5) with (6) produce in this case the usual gauge fixing 
term and the FP action. The gauge parameter $p$ could be 
$\beta$ or also some parameter in $F_{\alpha_i}$. Actually 
the explicit form of $\Psi$ in (6) will not be relevant 
below. \\
The idea of ``extended'' BRS transformation consists in 
including the gauge-parameter $p$ in the BRS operations,  
$ sp = z, \; sz = 0\,$  or, equivalently, to introduce a 
new (nilpotent $\sigma^2 = 0$) operation
\begin{equation}
\sigma = z\, \frac{\partial}{\partial\, p}\, , 
%eqn:7
\end{equation}
because the (anticommuting) $z$ obeys $z^2 = 0$ trivially.\\
The extended generating functional, depending on $z$ is now
\begin{equation}
Z (j,k,z) = \int\, (d\, \phi_A) \exp\, i \, L\; ,
%eqn:8
\end{equation}
where $L$ differs from $L^{(0)}$ only by the 
replacement\footnote{This trick has been used successfully 
to also incorporate ``external'' (global) symmetries of the 
action in order to obtain e.g.\ consistency conditions for 
anomalies in a very compact manner \cite{kum90}.} 
\begin{equation}
L^{(0)}_{gf} \; = \; s\,\Psi \; \to \; L_{gf} = (s+\sigma)\, 
\Psi\; .
%eqn:9
\end{equation}
The derivation of the Slavnov-Taylor identity proceeds as 
usual by the BRS-transformation (4)  
inside the path integral (8). The only 
terms in $L$ affected are the source term involving $j_A$  
and the new term with $\sigma$ from (9):
\begin{equation}
0 =  \int (d \phi) \; [\, j_A \delta \lambda s \phi_A + 
\sigma \delta \lambda s \Psi\,] \; \exp i L \; .
%eqn:10
\end{equation}
In (10) we have tacitly assumed BRS-invariance of the 
measure $(d\phi)$. This holds for gauge theories based upon 
Lie-groups. But also in other cases a covariant measure can 
be constructed, introducing e.g.\ the famous factor $\sqrt[4]{-g}$ for 
diffeomorphism invariance for the path integral of matter 
fields 
in gravity \cite{fuji88}. The global parameter 
$\delta \lambda$ is (anti-)commuted to the left and dropped, 
$s\phi_A$ replaced by $i^{-1}\,\delta/\delta k_A$ and 
$s\Psi$ by $L^{(0)}_{gf}$ (cf.\ (5)). \\
Clearly a path-integral like (1) or (8) is only meaningful 
if some kind of regularization is implied which, however, we 
do not have to specify. We shall not assume that the 
renormalization (with counterterms added to $L$) has been 
performed as yet. Therefore, the entire dependence on $p$ 
still resides in $L^{(0)}_{gf}$ (and nowhere else in 
$L^{(0)}$). Thus in the second term of (10) $s \Psi$ may be 
replaced even by the full extended $L$ because $\sigma^2 = 0$. 
This term simply corresponds to the action of (7)  
on $Z$: 
\begin{equation}
z \, \frac{\partial Z}{\partial p} = (-1)^A \; 
j_A \; \frac{\delta Z}{\delta k_A}
%eqn:11
\end{equation}
Expanding $ Z = Z^{(0)} + z\, Z^{(1)}$ the part linear in 
$z$ of (11)
\begin{equation}
\frac{\partial Z^{(0)}}{\partial p} = (-1)^{A+1} \, j_A\, 
\frac{\delta Z^{(1)}}{\delta k_A}
%eqn:12
\end{equation}
yields  all the information needed below. At $j_A = 0$ from (12) immediately 
follows  the gauge-independence of the vacuum loop 
contribution $Z^{(0)} (j=k=0) \equiv Z^{(0)} (0)$.\\
For the Green functions with $N$ external legs 
\begin{equation}
G_{A_1 \, \dots \; A_N} = \left.
\frac{(-i)^N}{Z^{(0)}\, (0)} \; 
\frac{\delta^N Z^{(0)}}{\delta\, j_{A_1} \, \dots \; \delta\, 
j_{A_N}} \; 
\right|_{j= k = 0}
%eqn:13
\end{equation}
Multiplication of (12) with the gauge-independent factor $[ 
Z^{(0)} (0)]^{-1}$ and $N$-fold differentiation as in (13) 
at $j=k=0$ leads to the identity
\begin{equation}
\frac{\partial\, G_{A_1\, \dots \, A_N}}{\partial p} \; = \; 
\sum\limits_{\ell=1}^N\;(-1)^{1 + \sum_{\nu=\ell}^{\nu=N} A_\nu}\, 
G^{(1)}_{A_1 \dots \ul{A}_\ell \dots A_N}\, ,
%eqn:14
\end{equation}
where $G^{(1)}$ is defined exactly like $G$, but with one of 
the $j$-legs replaced by a $k$-leg:
\begin{equation}
G^{(1)}_{A_1 \dots \ul{A}_\ell \dots A_N}\; = \; 
\frac{(-i)^N}{Z^{(0)} (0)}\; 
\frac{\delta^N \, Z^{(1)}}{\delta j_{A_1} \dots \delta 
k_{A_\ell} \dots \delta j_{A_N}}
%eqn:15
\end{equation}
The special case of the propagator ($N = 2$ in (14)) yields
\begin{eqnarray}
\frac{\partial\, G_{AB}}{\partial p} & = & 
(-1)^{1+A+B} \; G^{(1)}_{\ul{A}\, B} + (-1)^{1+B} \, 
G^{(1)}_{A\, \ul{B}}\nonumber\\
& = & - G^{(1)}_{\ul{A}\, B} - (-1)^B G^{(1)}_{A\, \ul{B}}\, , 
%eqn:16
\end{eqnarray}
where in the second line the fact has been used that in 
$G_{AB}$ on the l.h.s.\ $A$ and 
$B$ are either both commuting or both anticommuting  
 ($\sum_{\nu=1}^N A_\nu = 0 \to A+B=0$).

\section{Mass-Shell, Polarizations}

All ingredients for the definition of the S-matrix are 
contained in the Green functions (13). We assume that 
external lines of the S-matrix element are determined by 
``physical particles on shell'' which are present if the Feynman 
amplitude for the Fourier transform $\widetilde{G}_{AB} (k)$ of 
the propagator $G_{AB}$ for a certain value of the 
four-momentum $k$ at $k^2 - m^2 = \mu \sim 0$ possesses a pole of 
the type 
\begin{equation}
\widetilde{G}_{AB} (k) 
\quad\raisebox{-0.2cm}{$ = \atop { 
\scriptscriptstyle\mu\; \sim \; 0}$} \quad \frac{g_{AB}}{Z\,\mu}
%eqn:17
\end{equation}
with the sum over spin states yielding the polarization 
tensor 
\begin{equation}
g_{AB} = \sum\limits_{(s)}\, \overset{(s)}{e}\!{}_A \, 
\overset{(s)}{e}{}^\ast_B
%eqn:18
\end{equation}
determined by means of an orthonormalized (finite 
dimensional) basis
\begin{equation}
\overset{(s)}{e}{\!}_A \, \overset{(t)}{e_A^\ast} = \delta_{(st)}
%eqn:19
\end{equation}
of polarization vectors $e_A$ which is not unique. The 
transformation 
\begin{equation}
\delta \overset{(r)}{e}{\!}_A \; = \; \delta\, H_{AB}\, 
\overset{(r)}{e}{\!}_B
%eqn:20
\end{equation}
with antihermitian $\delta H_{AB} = - \delta H_{BA}^\ast$ 
leaves (19) invariant. E.g.\ for Dirac fermions $g_{AB}$ 
yields the factor $(k\!\!\!/ + m)$, for gauge-bosons the 
projection operator $g_{AB}$ generically depends  on the 
gauge parameter $p$. \\
Of course, (17) may be the result of diagonalizing a  
mass-matrix. It simply means that in the inverse of 
$\widetilde{G}$, i.e.\ in the self-energy $\widetilde{\Gamma}_{AB}$ 
near $\mu \sim 0$ the eigenvectors of $\widetilde{\Gamma}_{AB}$,  
obeying $ \widetilde{\Gamma}_{AB}\, e_B \sim Z \mu e_A$, are taken as a 
basis, which spans the degenerate states (spin components) 
for a certain $\mu \sim 0$. We note that in any case for the 
regularized theory the new parameter $m^2$ in $\mu$ at this 
point of our argument may still contain constant (but gauge 
dependent) contributions from higher order graphs (in 
a perturbative expansion, as well as from a nonperturbative 
point of view). To ease notation in the following we will 
assume that $e_A = e_A^\ast$ is real for the physical states 
and thus $\delta H_{AB}$ in (20) is antisymmetric. The 
argument does not change for complex eigenvectors $e_A$. We 
also for simplicity from now assume that the external lines in 
$G_{A_1 \, \dots \, A_N}$ do not contain fermions. \\
Amputating the propagator $G_{AB}^{(0)} = G_{AB}$ from the 
j-lines (now with grading $A=B=0$) 
on the r.h.s.\ of (16) yields with the amputated 
``rest'' $Y$ from a $G_{\ul{A}B}^{(1)}$, resp.\ $G_{A\ul{B}}^{(1)}$
\begin{equation}
\frac{\partial G_{AB}}{\partial p} = - Y_{\ul{A} C} \, 
G_{CB} - G_{AC} \, Y_{C\ul{B}} \, .
%eqn:21
\end{equation}
Near $\mu \sim 0$ after Fourier transformation ($G \to \widetilde{G}, 
\quad \widetilde{Y}\,\vert_{\mu=0}\to y$) this relation becomes
\begin{equation}
\frac{\partial}{\partial p}\, \left( \frac{g_{AB}}{Z \mu} 
\right) \;  = \; 
- y_{\ul{A}C}\; \frac{g_{CB}}{Z \mu} - \frac{g_{AC}}{Z 
\mu}\; y_{C\ul{B}} \; .
%eqn:22
\end{equation}
As usual in such arguments in this step we have made the 
assumption that the mass shell $\mu = 0$ is not accidentally 
degenerate with, say, the (gauge-dependent) mass shell of a 
Higgs-ghost (as in the  'tHooft gauge in  tree 
approximation) or of a FP ghost.\\
The absence of a second order pole $\mu^{-2}$ on the r.h.s.\ 
of (22) implies the gauge independence $\partial \mu / 
\partial p = 0$ of $\mu$ and hence of the mass parameter 
$m^2$. The terms of ${\cal O}\, (\mu^{-1})$ lead to the 
relation
\begin{eqnarray}
%eqn:23
X_{AB} \; & = & \; - X_{BA}\; ,\\
X_{AB} & = & \sum\limits_{(r)} \, 
\left(\frac{\overset{(r)}{e}{\!}_A}{2\, Z}\; \frac{\partial 
Z}{\partial\, p} - y_{\ul{A}C}\, \overset{(r)}{e}{\!}_C - 
\frac{\partial \overset{(r)}{e}{\!}_A}{\partial p} \, \right) \; 
\overset{(r)}{e}{\!}_B\; .
%eqn:24
\end{eqnarray}

\section{S-Matrix}

The S-matrix element for $N$ physical external particles 
(on-shell in the sense of (17)) is obtained from the 
Fouriertransform $\widetilde{G}_{A_1 \, \dots \; A_N}$ of the 
Green-function (13) after amputation of the propagators in 
each external line, multiplying each line with its proper 
polarization vector $\overset{(s_i)}{e}{\!}_{A_i}$ with spin state 
$(s_i)$. Furthermore the external lines acquire a 
renormalization factor $\sqrt{Z_i}$. Taking the mass-shell 
limit $\mu_i = k_i^2 - m^2_i \to 0$ yields 
\begin{equation}
S = \prod\limits_{i=1}^N \; \lim_{\mu_i\to 0} \; \left[ 
\; \mu_i \, \sqrt{Z_i} \, \overset{(s_i)}{e}{\!\!}_{A_i} \, \right] \; 
\widetilde{G}_{A_1 \, \dots \, A_N}\; .
%eqn:25
\end{equation}
An overall factor like $\prod\limits_i \, \left[\, (2\pi )^3\, 2 
(m^2_i + \vec{k}_i^2 ) \, \right]^{-1/2}$  
from the external lines is irrelevant (cf.\ the 
gauge-independence of $m_i$!) and has been dropped. 
Differentiation of $S$ with respect to $p$ leads to
\begin{equation}
\frac{\partial S}{\partial p} = \prod\limits_{i=1}^N \; 
\lim_{\mu_i \to 0}\; \left[ \mu_i \sqrt{Z_i} \, \right] \; 
\sum\limits_{\ell=1}^N \, ( A_\ell + B_\ell + C_\ell )\; ,
%eqn:26
\end{equation}
where 
\begin{equation}
A_\ell  = \frac{1}{2\, Z_\ell} \; 
\frac{\partial Z_\ell}{\partial p}\, 
\overset{(s_1)}{e}{\!\!}_{A_1}\, \dots \, 
\overset{(s_N)}{e}{\!\!}_{A_N}\,\widetilde{G}_{A_1\, \dots \, 
A_N}\; ,
%eqn:27
\end{equation}
\begin{equation}
B_\ell  =  \overset{(s_1)}{e}{\!\!}_{A_1} \, \dots \, 
\frac{\partial\, \overset{(s_\ell)}{e}{\!\!}_{A_\ell} }{\partial\, 
p} \; \dots\,  
\overset{(s_N)}{e}{\!\!}_{A_N}\,  
\widetilde{G}_{A_1\, \dots \, A_N}\; ,
%eqn:28
\end{equation}
\begin{equation}
C_\ell = - \overset{(s_1)}{e}{\!\!}_{A_1} \, \dots \, 
\overset{(s_N)}{e}{\!\!}_{A_N} \, \dots \, 
G^{(1)}_{A_1 \, \dots \, \ul{A}_\ell \, \dots \, A_N}\; .
%eqn:29
\end{equation}
Here $\partial \mu / \partial p = 0$, and in $C_\ell$ Eq.\ (14) 
for $\partial\, \widetilde{G}/\partial p$ (in our simplified case 
with commuting $A_1 \ldots A_N$)  has been used. In the 
limit $\mu_i \to 0$ the amputation of the relevant 
propagator terms (17) at all external lines in 
$\widetilde{G}_{A_1 \, \dots \, A_N}$ in $A_\ell$ of (27) 
yields
\begin{equation}
A_\ell = \frac{1}{2 Z_\ell}\; \frac{\partial 
Z_\ell}{\partial p} \; 
\widetilde{G}^{\, amp}_{A_1\, \dots \, A_N}\; 
\overset{(s_1)}{e}{\!\!}_{A_1} \, \dots \, 
\overset{(s_N)}{e}{\!\!}_{A_N} \; 
\prod\limits_{i=1}^N \; \frac{1}{\mu_i Z_i}\; ,
%eqn:30
\end{equation}
where the orthogonality relation (19) has been used. In 
$B_\ell$ of (28) the same procedure works for all lines, 
except for the one with factor $\partial e_{A_\ell}/\partial p$. 
In the latter we shift the derivative from 
$\overset{(s_\ell)}{e}{\!}_{A_\ell}$ to the propagator factor by 
(cf.\ the action of $\partial/\partial p$ on (19))
\begin{equation}
\frac{\partial\,\overset{(s_\ell)}{e}{\!\!}_{A_\ell}}{\partial\, p}\; 
\frac{\overset{(r)}{e}{\!}_{A_\ell} \overset{(r)}{e}{\!}_{B_\ell}
}{\mu_\ell\, Z_\ell} \; = \; 
- \overset{(s_\ell)}{e}{\!\!}_{A_\ell}\; 
\frac{\partial\,\overset{(r)}{e}{\!}_{A_\ell}}{\partial\, p}\;
\frac{\overset{(r)}{e}{\!}_{B_\ell}}{\mu_\ell\, Z_\ell}
%eqn:31
\end{equation}
Finally the analogous amputation at the poles of the 
propagators in (29) at the line with external source 
$k_{\ul{A}_\ell}$ produces a factor with $y$ as defined in 
(22):
\begin{equation}
C_\ell = - \prod\limits_{i=1}^N \, \left( \frac{1}{\mu_i\, 
Z_i} \right) \, \overset{(s_1)}{e}{\!\!}_{A_1}\, 
\overset{(s_1)}{e}{\!}_{B_1} \, \dots \, 
\overset{(s_\ell)}{y}{\!\!}_{\ul{A}_\ell\, C_\ell}\, 
\overset{(r)}{e}_{C_\ell}\,\overset{(r)}{e}_{B_\ell} \,\dots \,  
\overset{(s_N)}{e}{\!\!}_{A_N}\, \overset{(s_N)}{e}{\!\!}_{B_N}\; 
\widetilde{G}^{\, amp}_{B_1 \, \dots \, B_N}
%eqn:32
\end{equation}
Double spin indices $(s_i), (r)$ are being summed everywhere. It should be 
emphasized that in this way the $G^{(1)}$-amplitude reduces 
to its pole contribution $\overset{(\ell)}{y}$ and that (32) 
(as (26) and (28)) becomes proportional to the 
same amputated ordinary Green function $\widetilde{G}^{\, 
amp}_{A_1 \, \dots\dots\, A_N}$. \\
If we formally extract the special factor 
$\overset{(s_\ell)}{e}{\!\!}_{B_\ell}$ also in (27) by 
\begin{equation}
\widetilde{G}^{\, amp}_{A_1 \dots A_N}\, \overset{(s_1)}{e}{\!\!}_{A_1} \, 
\dots \, \overset{(s_N)}{e}{\!\!}_{A_N} \; = \; 
\widetilde{G}^{\, amp}_{A_1 \dots B_\ell \dots A_N}\, 
\overset{(r)}{e}{\!}_{B_\ell} \, \overset{(r)}{e}{\!}_{A_\ell} \, 
\dots\, \overset{(s_N)}{e}{\!\!}_{A_N} \; ,
%eqn:33
\end{equation}
adding the contributions (27), (28), (29) one finds that 
$\overset{(r)}{e}{\!}_{B_\ell}$ is multiplied just by the 
expression $\overset{(\ell)}{X}_{A_\ell B_\ell}$, as 
introduced in (23), applied to the line $A_\ell$. Apart from 
that the factors $\mu_i$ cancel so that the limit in (25) is 
a finite (nonvanishing) term proportional to  $\Pi_i 
Z_i^{-1/2}$. Collecting in the sum of (26) the expressions 
resulting in this manner from (27) with (33), (28) and (32), the 
{\em variation} of the S-matrix element with respect to 
$\delta p$ has the structure $\delta_1\, S 
\propto \; \dots \; \sum_\ell \, \dots \; 
(X_{A_l\, B_l}\, \delta p )\, e_{B_\ell}$. So far the basis 
$e_A$ has not been changed by varying $p$. If we take 
advantage of the freedom to redefine $\overset{(r)}{e}{\!}_A$ 
according to  (20) such  that $\delta_2 S \propto \, \dots 
\; \sum_\ell \, \dots \; (\delta \, H_{A_\ell\, B_\ell}) \, 
e_{B_\ell}$ with the (antisymmetric) $\delta H$ adjusted by 
the (antisymmetric!) $X_{AB}$ to $\delta H_{AB} = - X_{AB} 
\, \delta p$, we arrive at $( \delta_1 + \delta_2)\, S = 
\delta\, S = 0$, i.e.\ at the gauge independence of the 
S-matrix. For complex $e_A$ antisymmetry of $\delta H_{AB}$ 
and $X_{AB}$ is simply replaced by antihermiticity. Fermionic 
external lines just introduce a few minus signs. 

\section{Conclusion}

The gauge-parameter independence of the S-matrix requires a 
redefinition of the basis of polarization vectors attached 
to the external legs of the amputated Green function. Our  
proof is inherently nonperturbative. It refers to the 
regularized, but not renormalized amplitude. Therefore, it 
even applies to any finite perturbative order also for a 
nonrenormalizable theory. This seems important, because 
(even renormalizable) gauge theories at low energies may be 
part of a larger theory (e.g.\ string theory), where higher 
modes have been ``integrated out'' in the low energy regime, 
yielding effective couplings with (large) negative mass 
dimension. 
Our argument was based upon a gauge theory which, following 
the canonical procedure through extended Hamiltonian and 
gauge fixing fermion \cite{frad75} after integration of certain ghost 
fields and momenta etc. yields the path integral (2). We did 
not have to specify the gauge-fixing term further than to be 
BRS-exact. Therefore, a wide class of gauge theories is 
covered as well, with more complicated ghost-interactions 
than the ones in nonabelian gauge theories\cite{costa75} based upon 
Lie-groups. Our direct approach completely avoids proceeding 
through the gauge-dependence of 1pi-vertices, required in 
proofs\cite{GG00} which use Nielsen identities\cite{N75}. 
Especially for an S-matrix element with an arbitrary number of 
external legs, a proof based upon identities for 1pi vertices 
would appear very cumbersome.\\
As we work at the level of a regularized, but not yet 
renormalized S-matrix element, we are also not forced to deal 
with possibly delicate questions resulting from the gauge 
parameter dependence of counterterms.\\
A key point in the derivation of the basic 
identity (11) for Green functions has been the 
gauge-invariance of the measure in (8). Hence a deeper 
analysis would be required for a theory where no such 
measure is available. It may be conjectured that a theory,   
where the renormalizability is caused by  an anomaly may be 
outside the range of applicability of the present proof. 
Nevertheless, the present line of argument presumably is 
also useful in that context to exhibit an eventual 
gauge-dependence (and hence unphysicality) of the S-matrix.\\ 
All questions regarding the gauge-independence with external 
zero mass particles have been avoided by assuming (wherever 
necessary) some spontaneous symmetry breaking which provides 
a suitable regularization without destroying gauge 
invariance. Clearly the consideration of strictly mass-less 
external particles within the Bloch-Nordsieck or Lee-Nauenberg 
\cite{bloch37} mechanism deserves further study, possibly employing 
again some aspects of the technique described here.\\
A final remark concerns the possible extension of the present 
argument to S-matrix elements with strongly interacting external 
particles (hadrons). Probably also here a proof, generalizing the 
one known so far only for the axial gauge in nonabelian gauge 
theories\cite{WK80} seems to be conceivable.

\end{document}